\documentstyle[aps,pre,twocolumn,graphics,epsfig]{revtex}

\begin{document}
\title{SIMULATIONS OF DENSE GRANULAR FLOW:\\
DYNAMIC ARCHES AND SPIN ORGANIZATION}
\author{S. Luding(*), J. Duran, E. Cl\'ement, J. Rajchenbach}
\address{\center Laboratoire d'Acoustique et d'Optique de la Mati\`ere\\
Condens\'ee, URA 800 CNRS, Universit\'e Pierre et Marie Curie,\\
4 place Jussieu, 75252 Paris Cedex 05, France\\
(*) Institut f\"ur Computeranwendungen 1,\\
Pfaffenwaldring 27, 70569 Stuttgart, Germany\\
e-mail: lui@ica1.uni-stuttgart.de\\
~\\}

\maketitle

\abstract{We present a numerical model for a two dimensional
(2D) granular assembly, falling in a rectangular container
when the bottom is removed. We observe the
occurrence of cracks splitting the initial pile into pieces, like in 
experiments. We study in detail various mechanisms connected to 
the `discontinuous decompaction' of this granular material. 
In particular, we focus on the history of one single long range crack, 
from its origin at one side wall, until it breaks the assembly
into two pieces. This event is correlated to an
increase in the number of collisions, i.e. strong pressure, and to a 
momentum wave originated by one particle. Eventually,
strong friction reduces the falling velocity
such that the crack may open below the slow, high pressure `dynamic arch'.
Furthermore, we report the
presence of large, organized structures of the particles' angular
velocities in the dense parts of the granulate when the number of
collisions is large.\\
PACS: 46.10.+z, 05.60+w, 47.20.-k} 

\maketitle


\def\FFIGAA{{\center{\psfig{figure=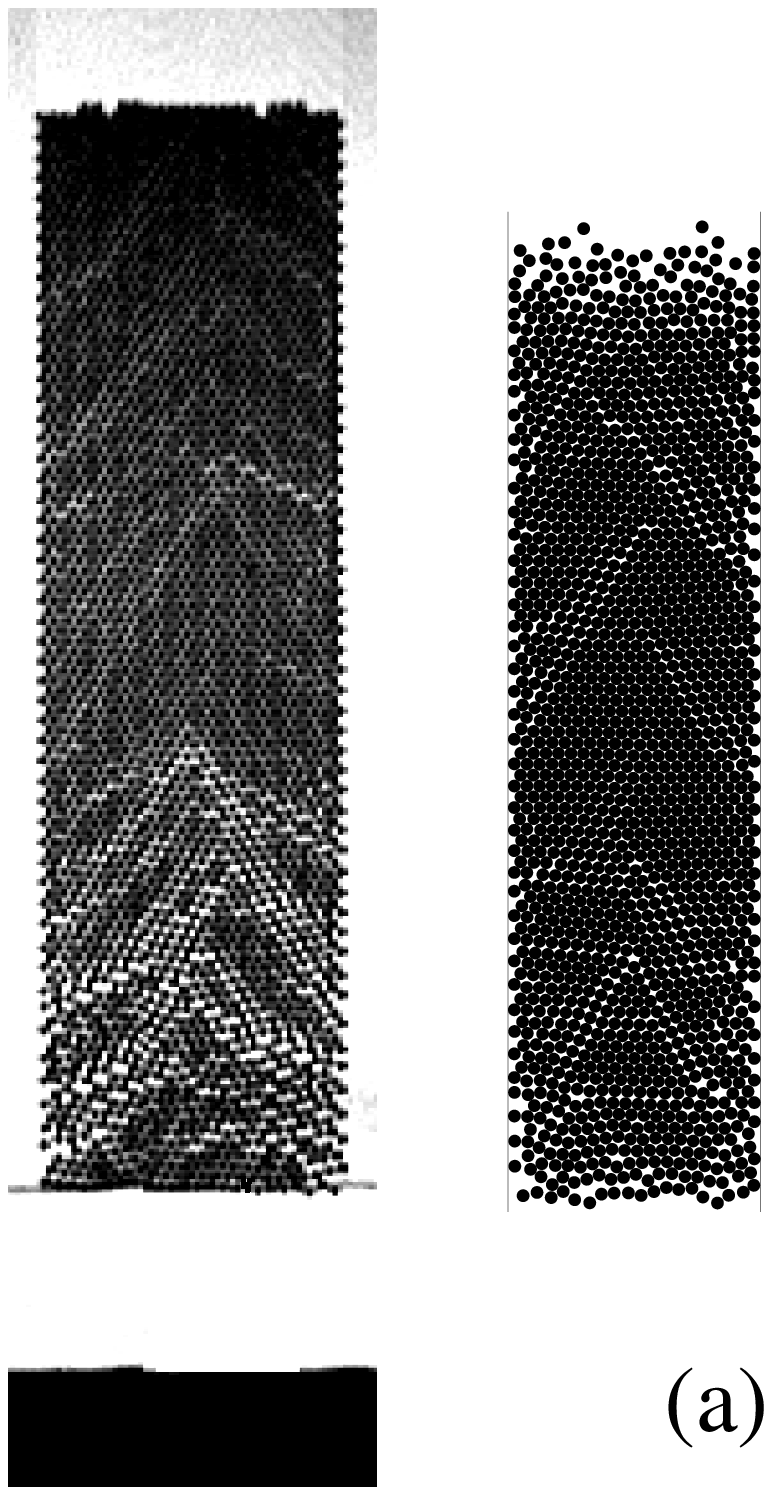,height=12cm}}}}
\def\FFIGA{
\begin{figure*}
\twocolumn[\psfig{figure=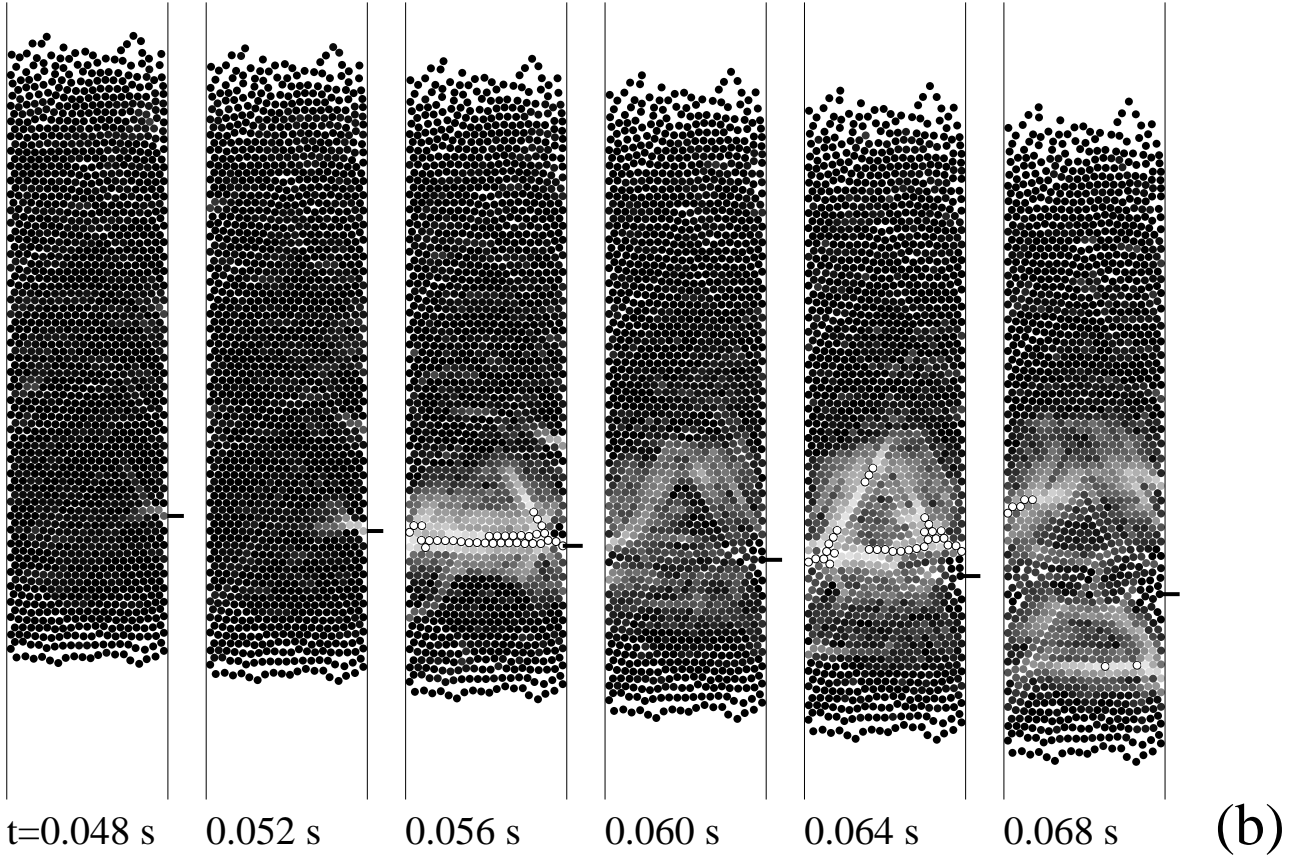,height=12.5cm}]
{\center{\psfig{figure=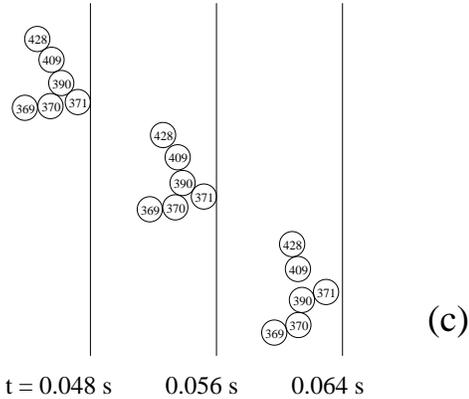,height=5.7cm}}}
\caption{
(a) Snapshots of a typical experiment (left) and a simulation 
(right) with $N$ =1562 particles,
at time $t$ = 0.06 s, in a pipe of width $L/d$ = 20.2. 
We used here $\epsilon$ = 0.96, $\epsilon_w$ = 0.92, 
$\mu$ = 0.5, $\mu_w$ = 1.0, and $\beta_0 = \beta_{0w}$ = 0.2. 
(b) Snapshots from a simulation with almost the same parameters as
in (a), but here $\epsilon$ = 0.90, $\epsilon_w$ = 0.90, 
and $\mu_w$ = 0.5. The greyscale indicates the
number of collisions per particle per millisecond. Black 
and white correspond to no collision or more than ten
collisions respectively. 
(c) Here only the selected particles \#369, \#370, \#371, 
\#390, \#409, and \#428 from (b) are plotted to illustrate
their motion. The vertical line indicates the right wall.
}
\label{fig:fig1}
\end{figure*}
}

\def\FFIGB{
\begin{figure}
{\center{\psfig{figure=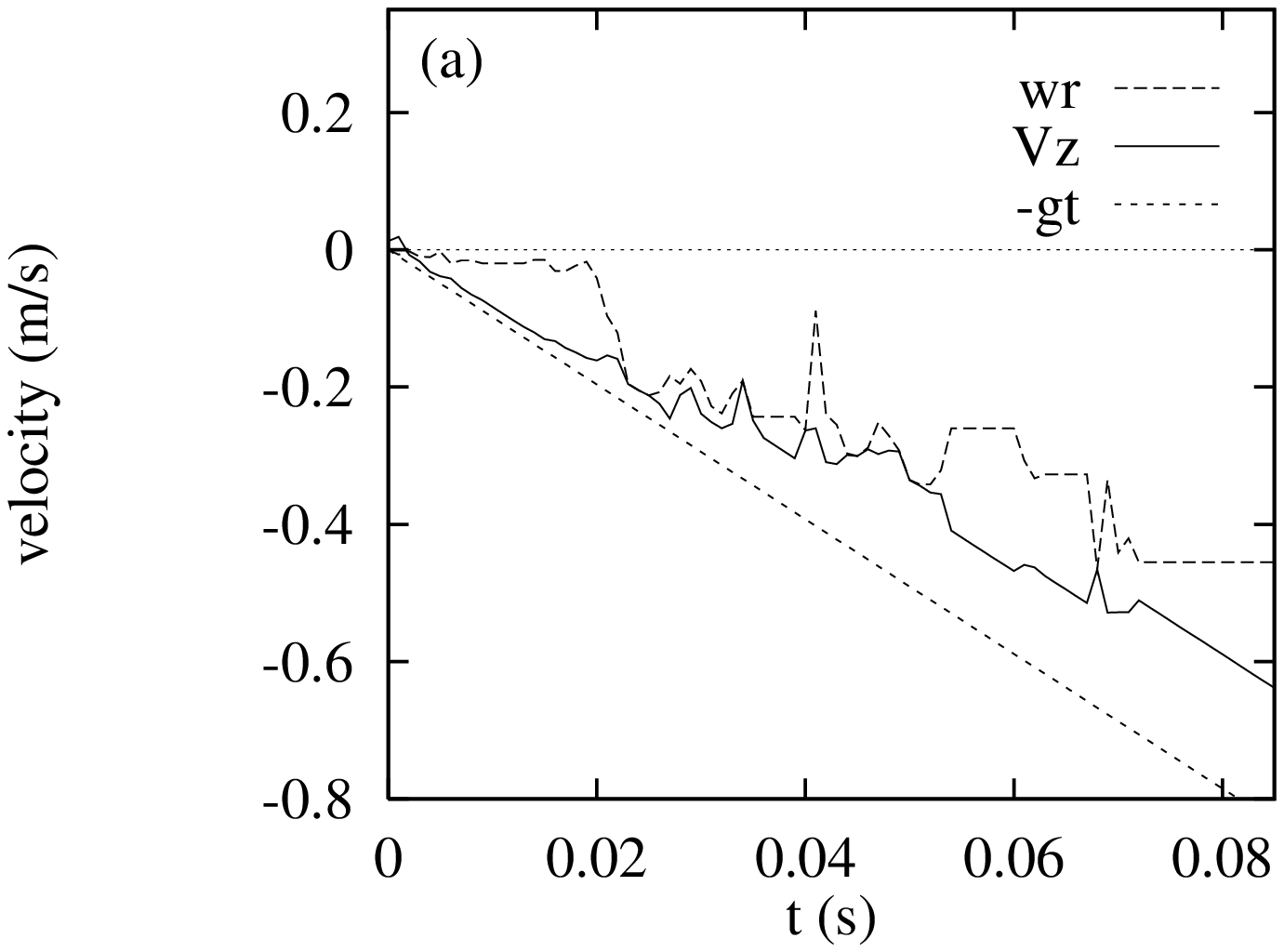,height=7cm}}}
{\center{\psfig{figure=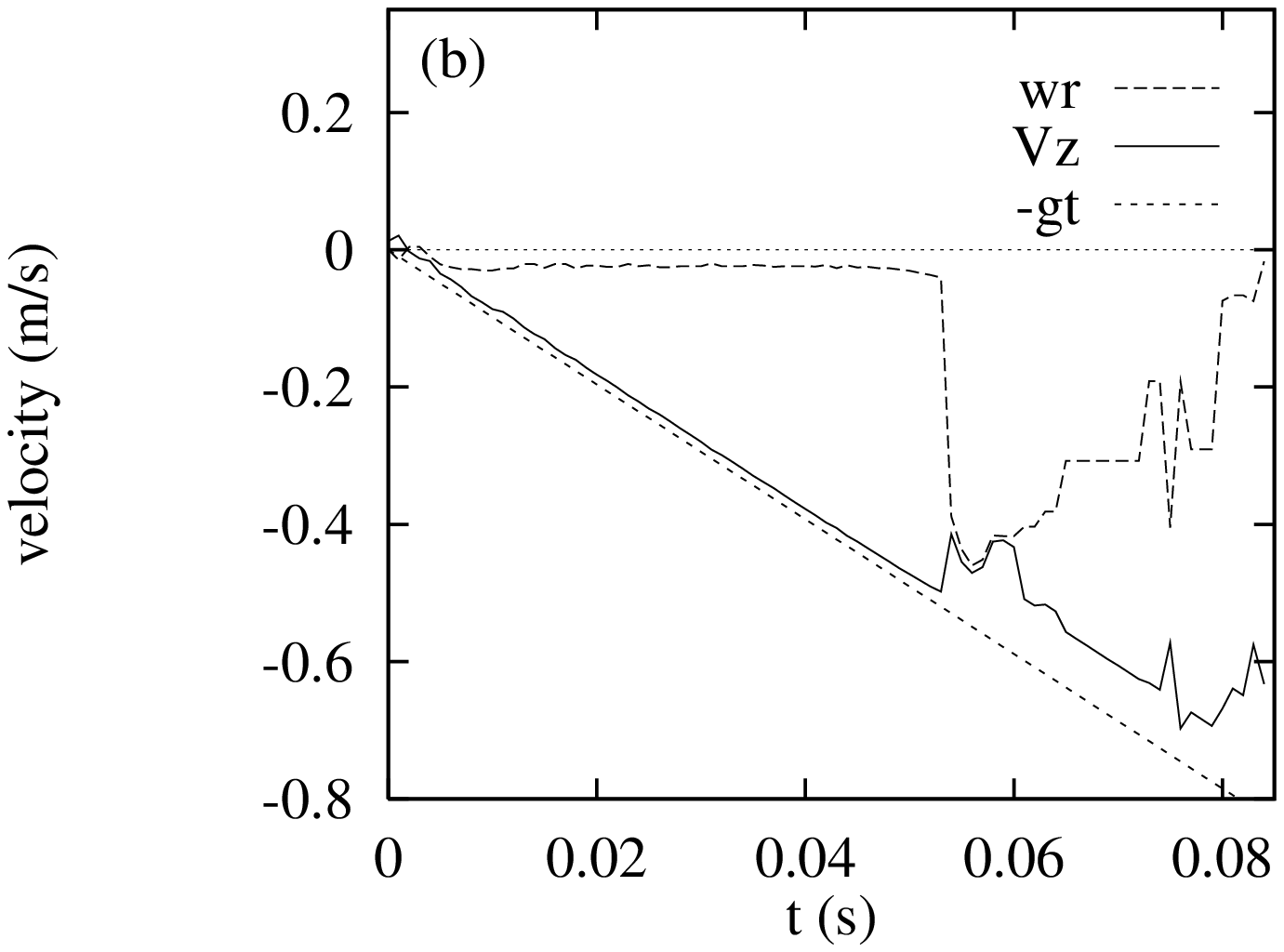,height=7cm}}}
\caption{
Plot of the angular velocity, $\omega r$, and of the 
linear, vertical velocity, $V_z$, of particle \#371 
versus time. The simulations are the same as in Fig.\
\protect\ref{fig:fig1}(a) and (b). The line $-gt$ 
corresponds to a freely falling particle.
}
\label{fig:fig2}
\end{figure}
}

\def\FFIGC{
\begin{figure}
{\center{\psfig{figure=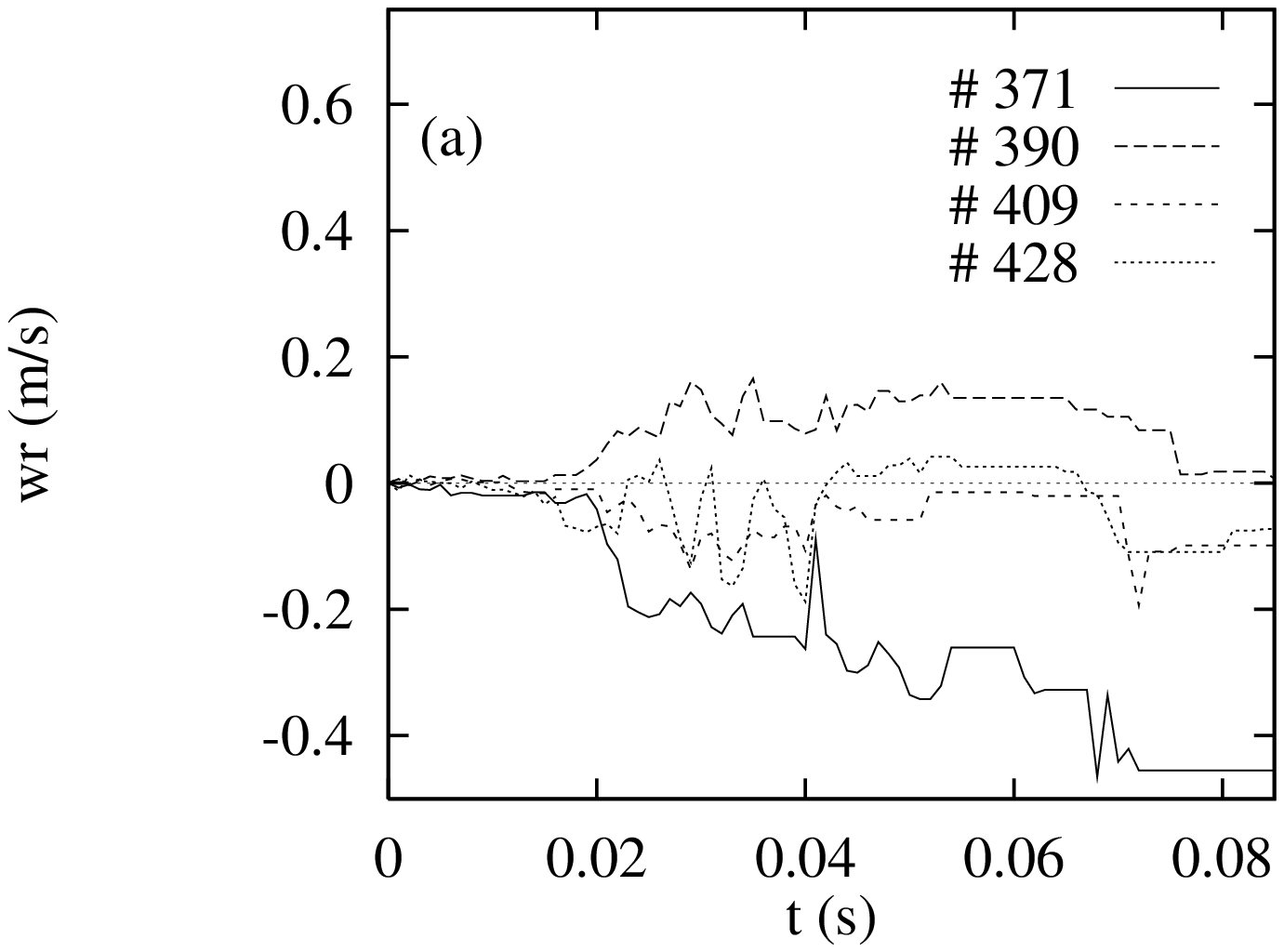,height=7cm}}}
{\center{\psfig{figure=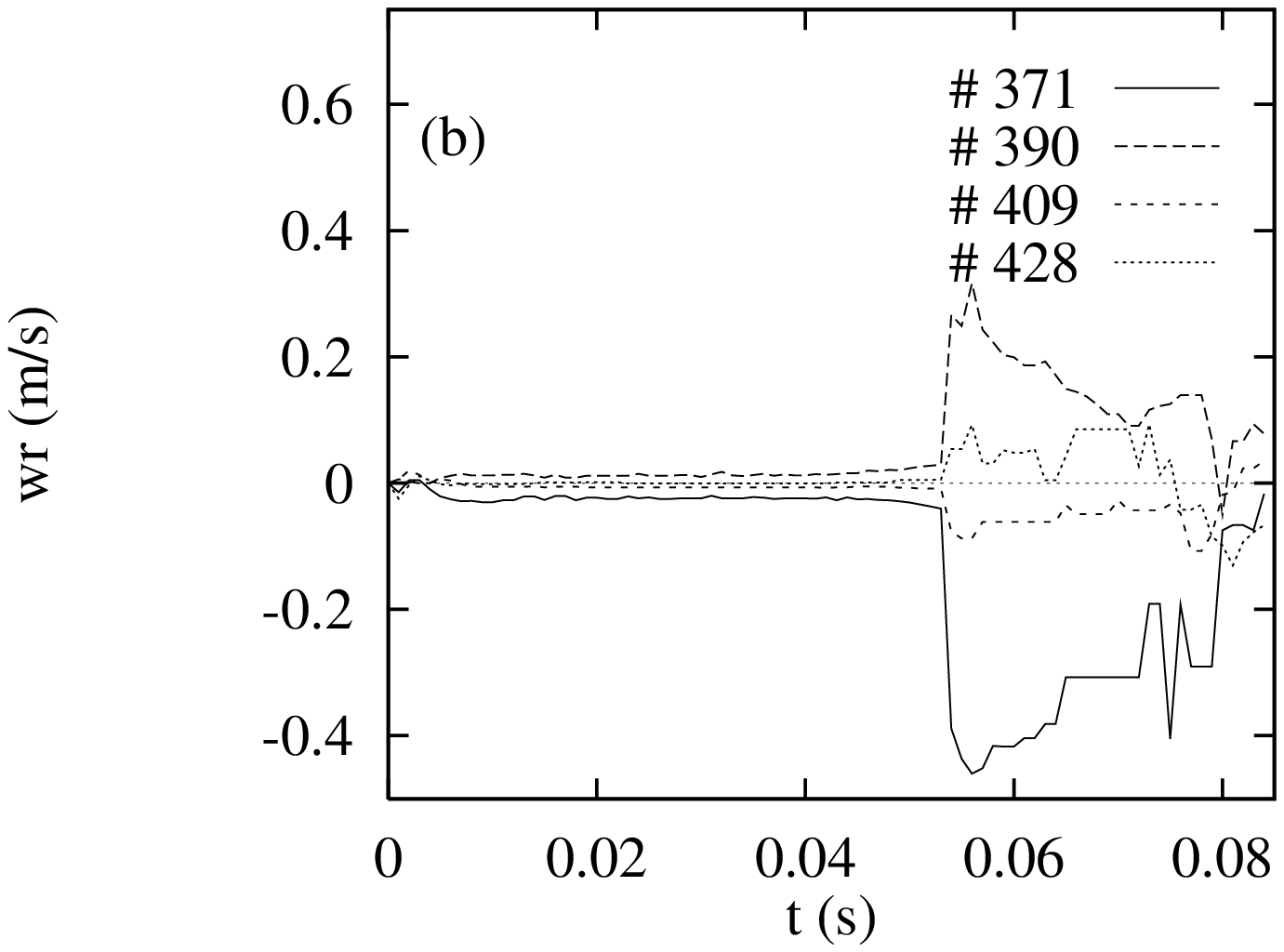,height=7cm}}}
\caption{
Plot of the angular velocities of the particles
\#371, \#390, \#409, and \#428 for the same simulations
as in Fig.\ \protect\ref{fig:fig1}(a) and
(b). These particles were initially arranged on
a line, tilted clockwise by 60 degrees from the horizontal, see
Fig.\ \protect\ref{fig:fig1}(c).
}
\label{fig:fig3}
\end{figure}
}

\def\FFIGD{
\begin{figure}
{\center{\psfig{figure=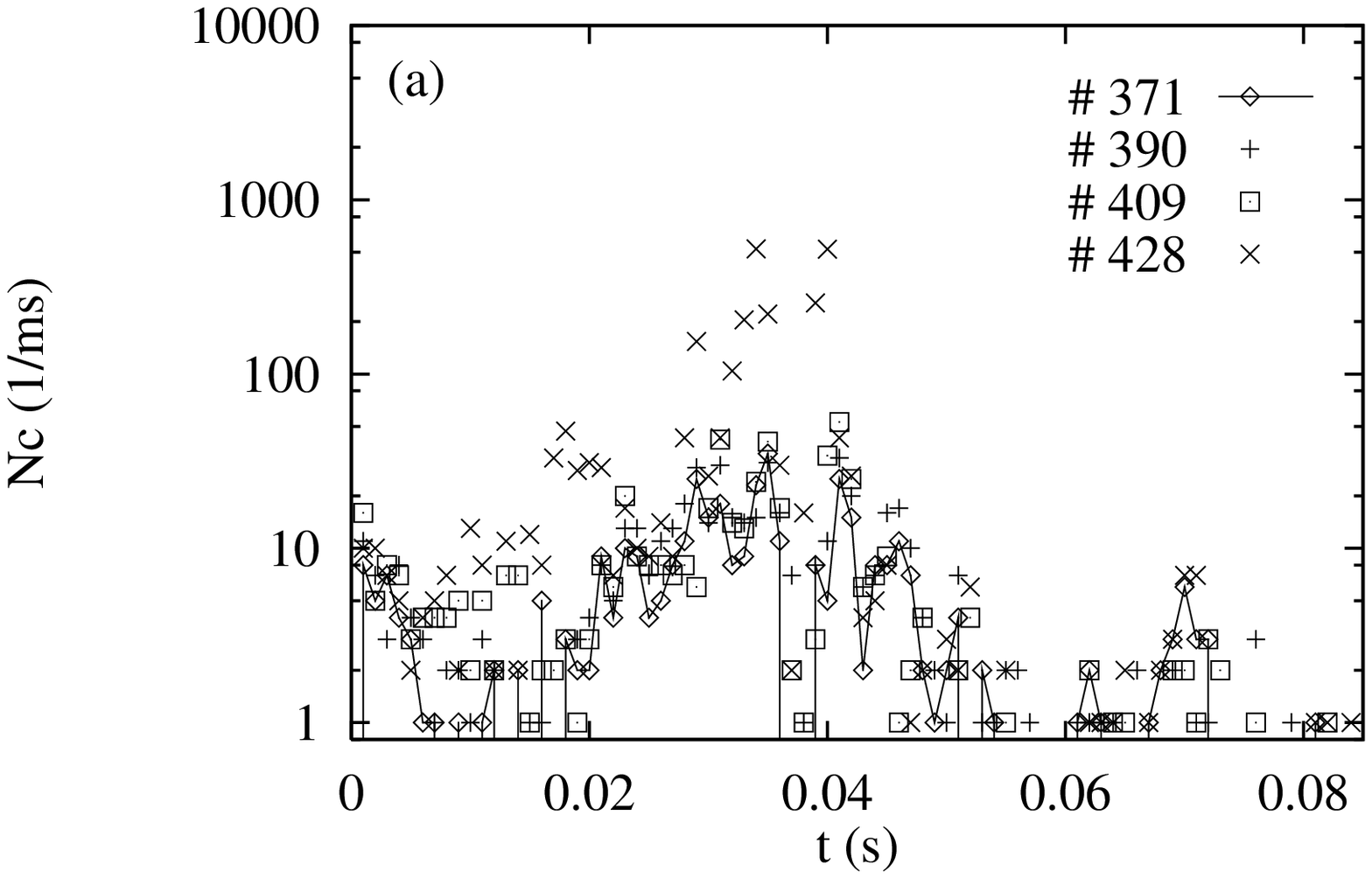,height=6cm}}}
{\center{\psfig{figure=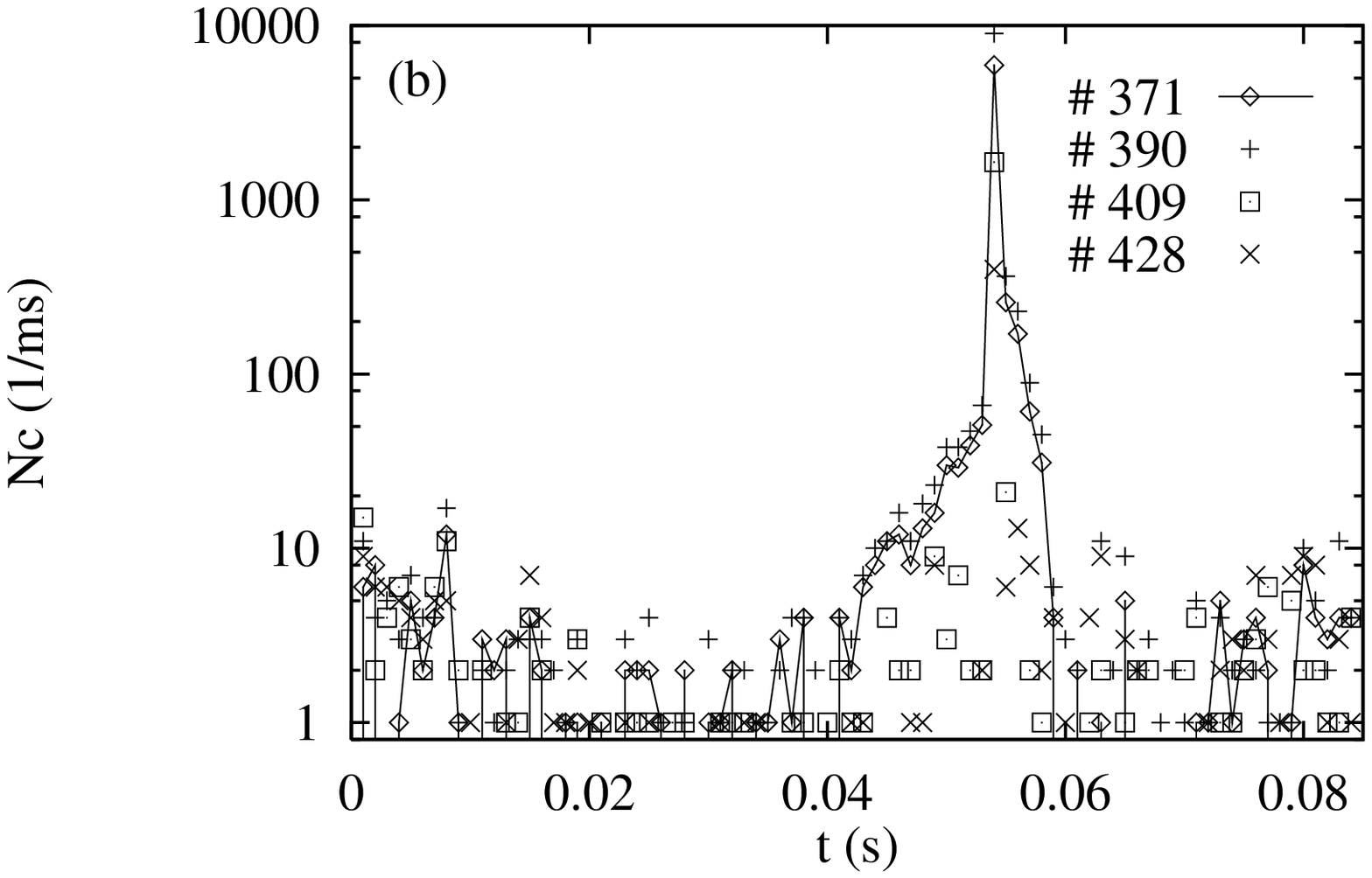,height=6cm}}}
\caption{
Log-lin plot of the number of collisions, $N_c$, per millisecond
(ms) as a function of simulation time, $t$. Plotted is $N_c$ for
the four particles noted in the figure. The data are from the 
same simulations as already presented in 
Fig.\ \protect\ref{fig:fig1} (a) and (b).
}
\label{fig:fig4}
\end{figure}
}

\def\FFIGE{
\begin{figure}
{\center{\psfig{figure=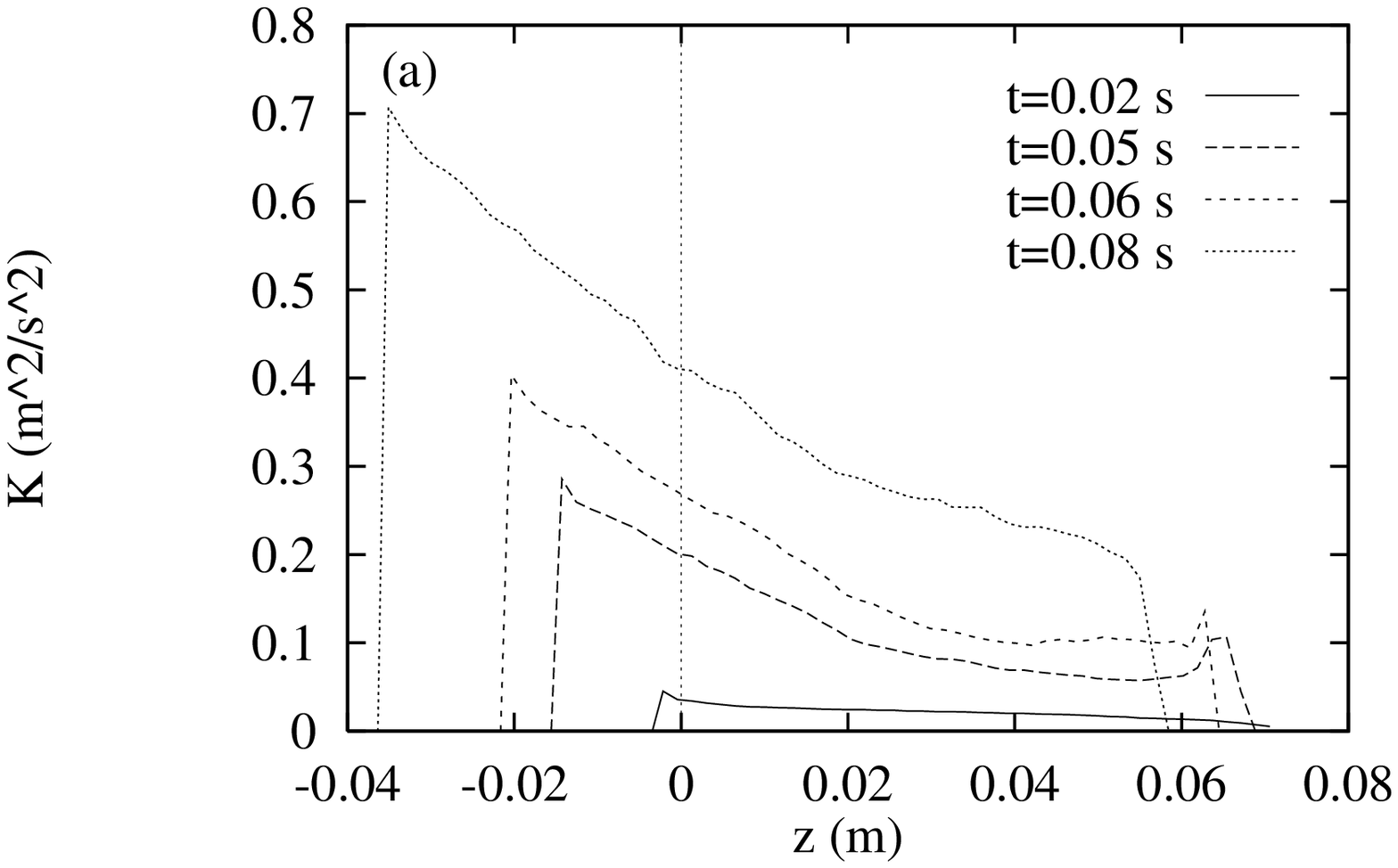,height=6cm}}}
{\center{\psfig{figure=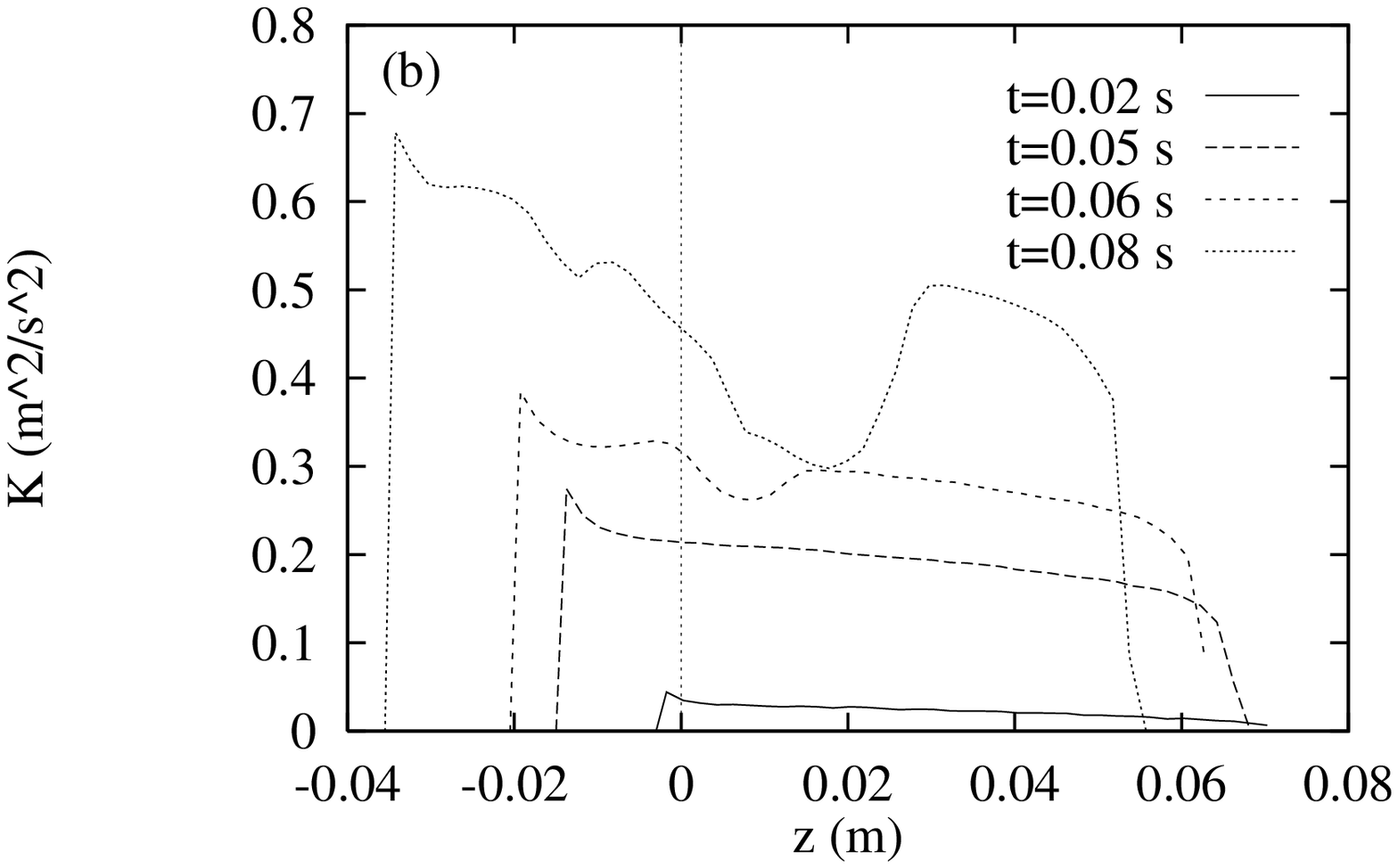,height=6cm}}}
\caption{
Plot of the kinetic energy $K$ as a function of height $z$
for different times for the same simulations as already presented in
Fig.\ \protect\ref{fig:fig1} (a) and (b).
}
\label{fig:fig5}
\end{figure}
}

\def\FFIGF{
\begin{figure}
{\center{\psfig{figure=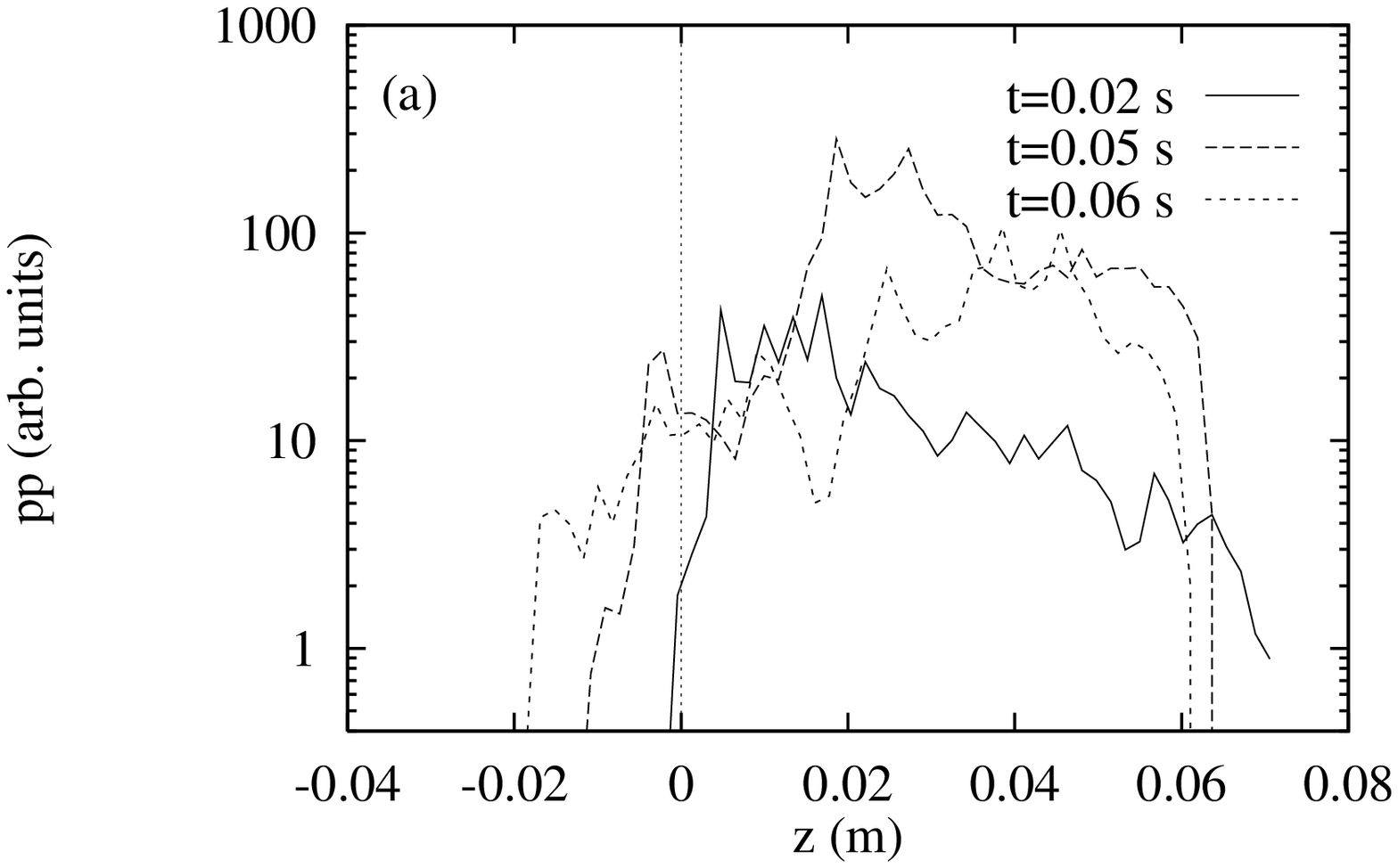,height=6cm}}}
{\center{\psfig{figure=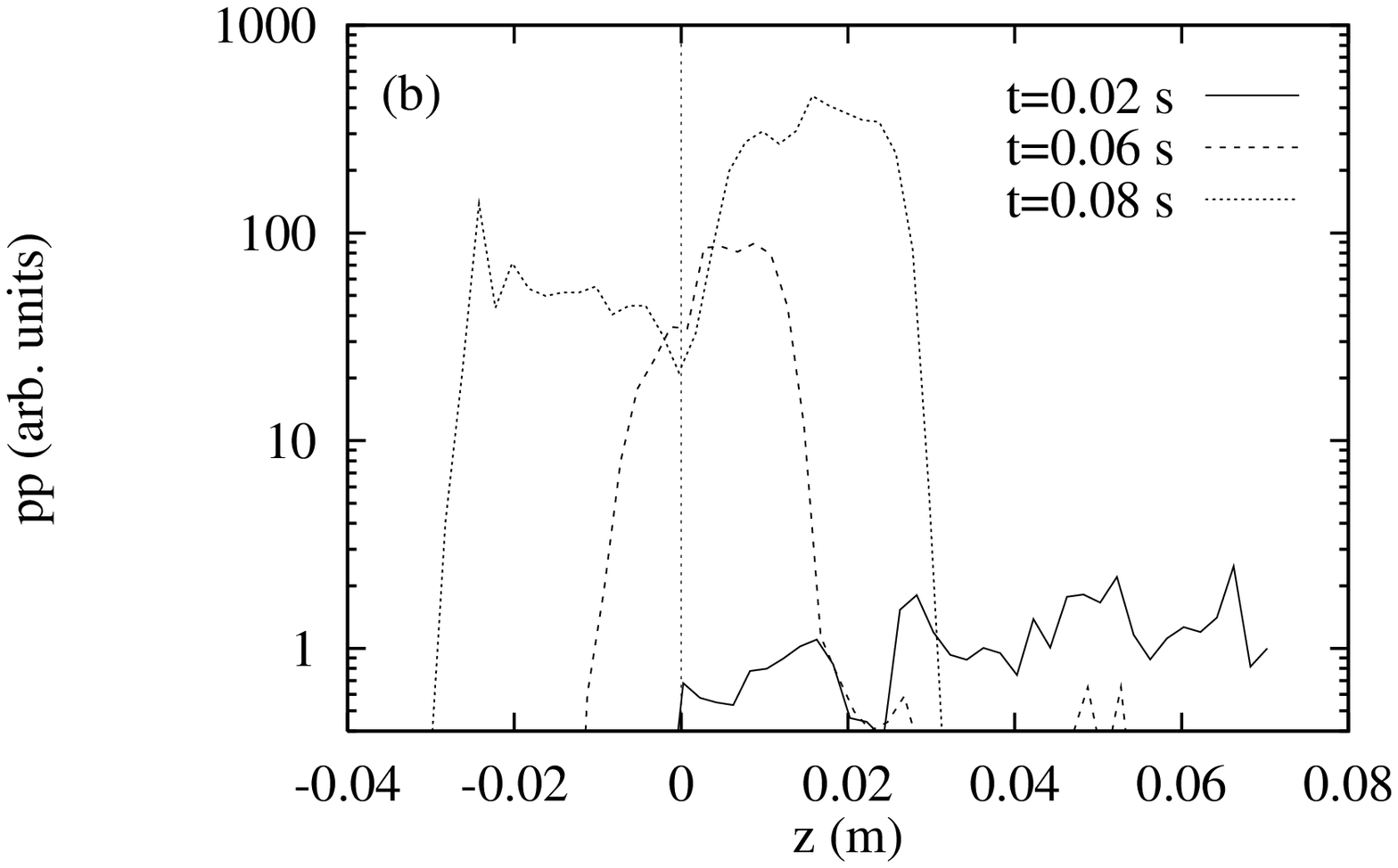,height=6cm}}}
\caption{
Log-lin plot of the pressure, $pp$, as a function of height $z$
for different times for the same simulations as already presented in
Fig.\ \protect\ref{fig:fig1} (a) and (b).
}
\label{fig:fig6}
\end{figure}
}

\def\FFIGG{
\begin{figure*}
{\center{\psfig{figure=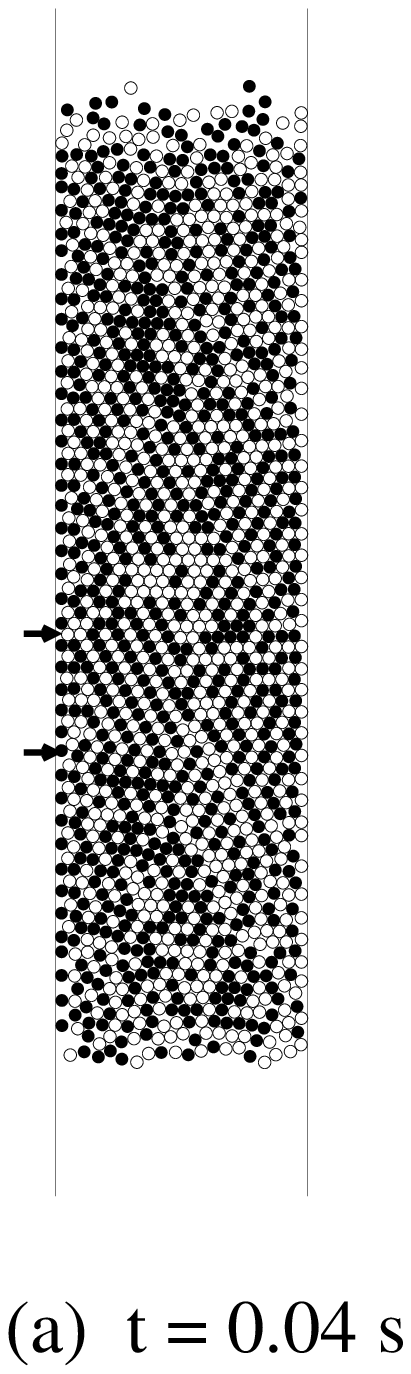,height=14cm}}}
\twocolumn[{\center \psfig{figure=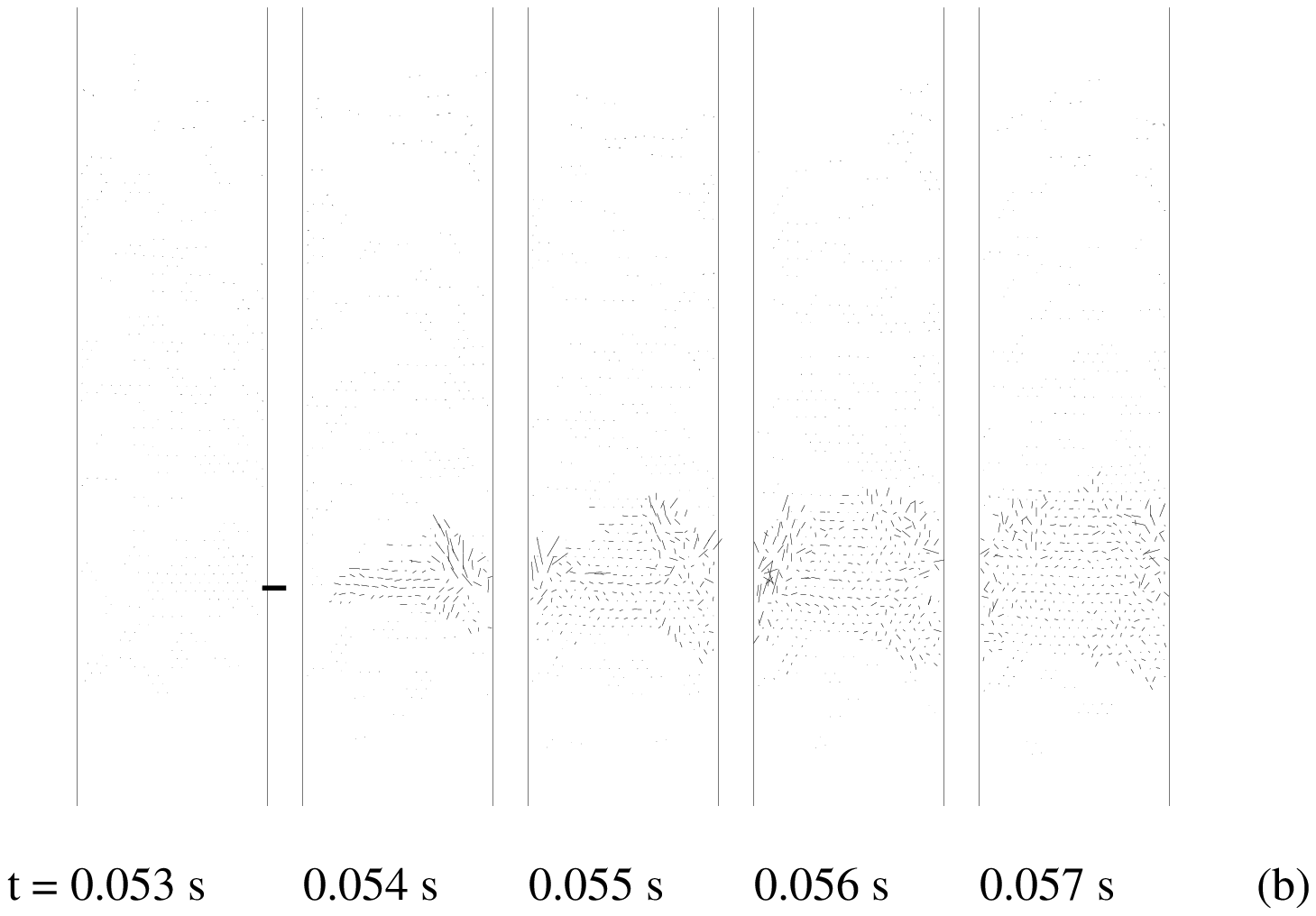,height=14cm}}]
\caption{
(a) Snapshots of the simulation from Fig.\ \protect\ref{fig:fig1}(a)
at different times. The greyscale indicates here the direction
of the angular velocity, i.e. black and grey correspond to 
clockwise and counterclockwise rotation respectively. 
(b) Snapshots from the simulation from Fig.\ \protect\ref{fig:fig1}(b)
at different times. The lines indicate for each particle the 
change in velocity due to collisions within the last 
millisecond, i.e. the last image.
}
\label{fig:fig7}
\end{figure*}
}

\section{Introduction}

The flow behavior of granular media in hoppers, pipes or chutes has received
increasing interest during the last years. For a review concerning
the physics of granular materials, see \cite{jaeger92,jaeger94} and refs.
therein.

The complete dynamical description of gravity driven flows is an open
problem and for geometries like hoppers or vertical pipes,
several basic phenomena, are still unexplained. In hoppers
intermittent clogging due to vault effects \cite{brown70},
density waves in the bulk \cite{baxter89} or 1/$f$ noise of the
outlet pressure, have been reported from experiments \cite{baxter93}. 
In a vertical pipe geometry, numerical simulations on
model systems with periodic boundary conditions \cite{poschel94,peng94} 
and also analytical studies \cite{lee94} show density waves. 
But so far, experimental evidence of density waves is
only found in situations where pneumatic effects, 
i.e. gas-particle interactions, are important \cite{herrmann95}.

In the rapid flow regime, kinetic theories \cite{jenkins83,lun84} describe
the behavior of the system introducing the granular temperature as a measure
of the velocity fluctuations. In quasistatic situations, also arching effects
and particle geometry get to be important. Under those conditions, kinetic
theories and also continuum soil mechanics do not completely describe all
the phenomena observed. The complicated particle-wall and
particle-particle interactions \cite{radjai95,radjai95b,radjai95c}, 
possible formation of stress chains in the granulate and also stress 
fluctuations \cite{pouliquen95,liu95} 
undoubtedly require further experimental, theoretical and simulation
work.

Following recent research on model granular systems 
\cite{pouliquen95,liu95,duran94,knight93,clement92,warr94,cantelaube95} 
we focus here
on the problem of a 2D pile made up of rather large spheres enclosed in a
rectangular container with transparent front and back walls, separated by
slightly more than one particle diameter. Recent observations of
approximately V-shaped {\it microcracks} in vertically vibrated sand-piles 
\cite{duran94} were complemented by recent experiments and simulations of
the discontinuous decompaction of a falling sandpile \cite{duran95}. From
experiments Duran et al. \cite{duran95} find the following basic features:
In a system with polished laterals walls, cracks are unlikely to appear
during the fall, i.e. the pile will accelerate continuously, with an
acceleration value depending on the aspect ratio of the pile
and on the friction with the walls. 
In a system with rather poorly polished walls (a surface 
roughness of more than 1 $\mu$m in size),
cracks occur frequently. A crack in the lower part of the
pile grows, whereas a crack in the upper part is unstable and will
eventually close. These experimental 
findings can be understood from a continuum approach based on a dynamical
adaptation of Janssen's model \cite{brown70,duran95}. 
Nevertheless, not all of
the discontinuous phenomena, such as the reasons for the cracks,
can be explained by such a continuum model. In previous
works, numerical simulations were used to parallel the experiments and to
analyze the falling pile for different material's parameters \cite
{duran95,luding95c}. 
Though the simulations are dynamic, in contrast to 
an experiment which starts from a static situation, a reasonable
phenomenological agreement was found. Furthermore, simulations were able to
correlate the existence of long range cracks to strong local
pressure on the walls. The increase in pressure was found to be
about one order of magnitude \cite{duran95}.
In this work we use the numerical model of Refs. \cite
{walton92,duran95,luding95c,luding95b} and investigate in detail
how a crack occurs. In particular, we follow one specific crack
and try to extract the generic features that could be
relevant for a more involved theoretical description of the 
behavior of granular materials. 

We briefly discuss the simulation method in section \ref{simulation}
and present our results in section \ref{results} which consists of
three main parts: Firstly, we present stick-slip
behavior and a local organization of the spins in subsection
\ref{sec:stick}. Secondly, we describe in how far the number of
collisions, the kinetic energy and the pressure are connected
in \ref{sec:ncoll} and finally we present long range organization
and momentum waves in subsection \ref{sec:order}. We summarize and
conclude in section \ref{conclusion}.

\section{The Simulation Method}

\label{simulation}

Our simulation model is an event driven (ED) method \cite
{walton92,duran95,luding95b,lubachevsky91,luding94c,foerster94} 
based upon the following
considerations: Particles undergo a parabolic flight in the gravitational
field until an event occurs. An event may be the collision of two particles
or the collision of one particle with a wall. Particles are hard spheres and
interact instantaneously; dissipation and friction are only active on contact.
Thus we calculate the momentum change using a model that is consistent
with experimental measurements \cite{foerster94}. From the change of 
momentum
we compute the particles' velocities after a contact from the velocities
just before contact. We account for energy loss in normal direction, as for
example permanent deformations, introducing the coefficient of normal
restitution, $\epsilon $. The roughness of surfaces and the connected energy
dissipation, is described by the coefficient of friction, $\mu$, and the
coefficient of maximum tangential restitution $\beta _0$. 
For interactions between
particles and walls we use an index $w$, e.g. $\mu_w$. Due to the
instantaneous contacts, i.e. the zero contact time, one may observe the
so-called `inelastic collapse' in the case of strong dissipation. For a
discussion of this effect see McNamara and Young \cite{mcnamara95} and refs.
therein. Despite this problem, 
we use the hard sphere ED simulations as one possible approach,
since also the widely used soft sphere molecular dynamics (MD) simulations
may lead to complications like the `detachment-effect' \cite
{luding94b,luding94d,luding95a} or the so-called 
`brake-failure' for rapid flow along rough walls \cite{schaefer95}. 
Recent simulations of the model system, described in this study, using an 
alternative simulation method, the so called 'contact dynamics' (CD)
\cite{radjai95c}, also lead to cracks \cite{moreau96}. The CD
method has a fixed time-step, in contrast to ED where the time-step
is determined by the time of the next event.

From the momentum conservation laws in linear and angular direction, from
energy conservation, and from Coulomb's law we get the change of 
linear momentum of particle 1 as a
function of $\epsilon$, $\mu$, and $\beta_0$ \cite{luding95b}: 
\begin{equation}  \label{eqdp}
\Delta \vec{P}=-m_{12}(1+\epsilon)\vec{v}_c^{(n)}- {\frac 27}%
m_{12}(1+\beta) \vec{v}_c^{(t)},
\end{equation}
with the reduced mass $m_{12} = m_1 m_2 / (m_1 + m_2)$. For particle-wall
interaction, we set $m_2 = \infty$ such that $m_{12} = m_1$. 
(n) and (t) indicate the normal and tangential components of 
the relative velocity of the contact points 
\begin{equation}  \label{eqvc}
\vec{v}_c = \vec{v}_1-\vec{v}_2- \left( {{\frac{d_1}2}\vec{\omega}_1+{\frac{%
d_2}2}\vec{\omega}_2} \right) \times \vec{n},
\end{equation}
with $\vec{v}_i$ and $\vec{\omega}_i$ being the linear and angular
velocities of particle $i$ just before collision. 
$d_i$ is the diameter of particle $i$ and the unit vector in
normal direction is here definded as $\vec n = (\vec r_1 - \vec r_2)
/ | \vec r_1 - \vec r_2 |$. Paralleling $\epsilon$, the
(constant) coefficient of normal restitution we have $\beta$, the
coefficient of tangential restitution 
\begin{equation}  \label{eqbeta}
\beta = \min \left [ \beta_0, \beta_1 \right ].
\end{equation}
$\beta_0$ is the coefficient of maximum tangential restitution, 
$-1 \le \beta_0 \le 1$, and accounts for energy conservation and
for the elasticity of the material \cite{foerster94}. 
$\beta_1$ is determined using Coulomb's law
such that for solid spheres $\beta_1 = - 1 - (7/2) \mu (1+\epsilon) 
\cot \gamma$ with the collision angle $\pi/2 < \gamma \le \pi$ 
\cite{luding95b}. Here, we simplifed the tangential
contacts in the sense that exclusively Coulomb-type interactions, i.e. $%
\Delta P^{(t)}$ is limited by $\mu \Delta P^{(n)}$, or sticking contacts
with the maximum tangential resitution $\beta_0$ are allowed. 
Sticking corresponds thus to the case of low tangential velocity
whilst the Coulomb case corresponds to sliding, i.e. a 
comparatively large tangential velocity. For a detailed
discussion of the interaction model used see Refs.\ \cite
{walton92,luding95b,foerster94}.

\section{Results}
\label{results}

Since we are interested in the falling motion of a compact array of
particles, we first prepare a convenient initial condition. Here, we use $%
N=1562$ particles of diameter $d=1$ mm in a box of width $L=20.2d$, and let
them relax for a time $t_r$ under elastic and smooth conditions until the
density and energy profiles do not change any longer. The choice of $L$
is quite arbitrary, however, we wanted to start with a triangular 
lattice with a lattice constant of $d (1+\Delta)$ and $\Delta$ to
be small but larger than zero, i.e. $\Delta = 0.01$. We tested several
height to width ratios $S = H/L$ of the system and found e.g. the same behavior
as in experiments, i.e. the larger $S$, the stronger are the effects
discussed in the following. The average velocity
of the resulting initial condition is $\overline{v}=\sqrt{<v^2>}\approx
0.05$m/s. 
\FFIGAA
\FFIGA 
The kinetic energy connected to $\overline{v}$ is comparable 
to the potential energy connected to the size of one particle.
Due to this rather low
kinetic energy, the array of particles is arranged in triangular order,
except for a few layers at the top. Some tests with different values of
$\overline{v}$ lead to similar results as long as $\overline{v}$ is
not too large. The larger $\overline{v}$, the more particles belong to 
the fluidized part of the system at the top and in the fluidized part
of the system we can not observe cracks. For lower values of
$\overline{v}$ we observed an increasing number of events per
unit time and thus increasing computational costs for the simulation.
At $t=0$ we remove the bottom, switch on
dissipation and friction and let the array fall. 

Performing simulations with different initial conditions and different sets
of material's parameters, we observe strong fluctuations in position and
shape of the cracks. However, the intensity or the probability of the cracks
seems to depend on the material's parameters rather than on the initial
conditions. The behavior of the system depends on 
friction and on dissipation as well: For weak friction we observe
only random cracks, which would occur even in a dense hard sphere gas without
any friction, simply due to random fluctuations and the internal pressure. 
With increasing friction, cracks may even span the whole system 
and sometimes be correlated to slip planes. Furthermore, we observe
cracks to occur more frequent for lower dissipation.

In Fig.\ \ref{fig:fig1}(a) we present typical snapshots of an experiment 
and of a simulation at time $t$ = 0.04 s. We observe long range cracks 
from both, experiments and simulations as well. For the experiment
we use a container of width $L/d$ = 24 with 103 layers of 
oxidized aluminum particles. 
The vertical 2D cell is made up of two glass windows for visualisation
and of two lateral walls made of plexiglass. The gap between the glass
windows is a little larger than the bead diameter, what leads to 
a small friction between particles and front/back walls, while
the friction with the side walls may be large. Different heights and
wall-materials were used and the results could be scaled 
with a
characteristic length $\xi = L / (2 K \mu_w)$, with the dimensionless
parameter $K$ which characterizes the conversion of vertical to
tangential stresses and the coefficient of friction $\mu_w$.
The scaling in the regime before cracks occur leads to the 
value $K \mu_w \approx 0.12$. 
For a more detailed description of the experimental
setup and data see Ref.\ \cite{duran95}.
In the simulation of Fig.\ \ref{fig:fig1} we have $N$ = 1562 and the 
parameters $L/d$ = 20.2, $\epsilon$ = 0.96, $\epsilon_w$ = 0.92,
$\mu$ = 0.5, $\mu_w$ = 1.0, and $\beta_0 = \beta_{0w}$ = 0.2.
We varied the coefficients of friction in the intervals
$0 \le \mu \le 1$ and $0 \le \mu_w \le 10$. Furthermore, we varied
the coefficients of restitution in the range $0.80 \le \epsilon or
\epsilon_p \le 0.98$. However, the occurrence of cracks is
quite independent of the parameters used, as long as the coefficients of
friction, $\mu$ and $\mu_w$ are sufficiently large. Furthermore,
cracks occur faster for stronger dissipation since a highly
dissipative block dissipates the initial energy faster.

For the above parameters,
we find - like in experiments - a large number of cracks, overlapping
and interferring. In the upper part of both, experiment and 
simulation, we sometimes observe cracks only on one side. 
In contrast to Fig.\ \ref{fig:fig1}(a) we present in Fig.\ \ref{fig:fig1}
(b) a specific simulation with only one strong crack, on which we focus
in more detail. This crack separates the system into a large upper
and a small lower part and is best visible in Fig.\ \ref{fig:fig1}(b)
at $t$ = 0.068~s. Here we use a reduced wall friction, i.e.
$\mu_w$ = $\mu $ = 0.5, and stronger dissipation, i.e.
$\epsilon_w = \epsilon = 0.9$, while all other parameters, 
including the initial configuration, are the same as for (a).

In order to distinguish consistently, we will refer to the simulation
with large $\mu_w$ as simulation (a), and to the simulation with
small $\mu_w$ as simulation (b) in the following. 

The important feature of the crack in Fig.\ \ref{fig:fig1}(b) is
that it seems to be connected to one single particle. We
indicate the vertical position of particle \#371 with a small bar 
in Fig.\ \ref {fig:fig1}(b). The
crack of simulation (b) is connected to an increase of the number of
collisions per particle, indicated by the greyscale on Fig.\ \ref{fig:fig1}%
(b). Black or white correspond to zero or more than twenty collisions
during the last millisecond respectively. Note that the increase in 
the number of collisions is here equivalent to an increase in pressure.
Already at time $t$ = 0.048~s, particle \#371 peforms more collisions 
than the average particle. The pressure around particle \#371
increases and at $t$ = 0.056~s a region in which the particles
perform a large number of collisions spans the whole width of the container. 
We call such an array of particles under high pressure `dynamic arch'.
At $t$ = 0.060~s the pressure decreases and a large crack opens
below particle \#371. At later times, we observe new pressure fluctuations 
in the array. In conclusion, a crack seems to begin at one point, 
i.e. one particle, where the pressure increases accidentally. 

Before we look in more detail at the behavior of particle \#371, 
we present for convenience, a picture of some selected particles 
around \#371 from (b), in Fig.\ \ref{fig:fig1}(c). 

\subsection{Evidence for Stick-Slip Behavior}

\label{sec:stick}

From Fig.\ \ref{fig:fig1}(b) we evidenced that a crack may
originate from one particle only. Now we are interested in the 
velocity of one specific particle during its fall. Following
the order of simulations (a) and (b) we firstly present the case
of large wall friction (a) and secondly the case of smaller wall friction
(b) in the following. 
Remember that particle \#371 is the origin of one single crack
in simulation (b), while its behavior is similar to the behavior 
of many others, close to the boundary, in simulation (a).

\protect{\FFIGB}
Due to gravity, the particle is accelerated downwards and after a collision with
the right wall it will presumably rotate counterclockwise. Therefore,
we compare the linear velocity, $V_z$, with the rotational velocity of the
surface, $\omega r$. In Fig.\ \ref{fig:fig2} 
we plot both, the linear vertical and the angular velocity of
particle \#371 as a function of time. The
horizontal line indicates zero velocity and the diagonal line indicates the
free fall velocity, $-gt$. The full curve gives the (negative) linear
velocity in vertical direction, $V_z$, while the rotational velocity of the
surface, $\omega r$, is given by the dashed curve. Negative $\omega r$ values
correspond to counterclockwise rotation and for $V_z = \omega r$ we have
the contact point of the particle at rest relative to the wall. The particle
adapts rotational and linear velocity, or in other words, the
contact point sticks. This event occurs when the two curves merge.
In Fig.\ \ref{fig:fig2}(a) we
observe a small angular velocity up to $t \approx$ 0.02 s when particle
\#371 first sticks. Note that in this simulation (a) \#371 can not be
identified as the initiator of one of the numerous cracks; 
however, it sticks and slips several times. An increase
in angular velocity goes ahead with a decrease of linear velocity due to
momentum conservation. At larger times we observe the angular velocity
decreasing, i.e. the particle slips, and short time after, sticks again.

\protect{\FFIGC}
Now we focus on simulation (b) and the behavior of the
particle, from which the crack started.
In Fig.\ \ref{fig:fig2}(b) we observe a small angular velocity up to $t
\approx$ 0.050~s. At $t$ = 0.055~s the velocities are adapted for some
0.005~s before the particle slips again. 
Since the pressure fluctuations are
visible in Fig\ \ref{fig:fig1}(b) already at $t$ = 0.048~s, 
we conclude that pressure fluctuations in the bulk lead 
to a sticking of a particle surface on the wall. 
This particle is slowed down and thus will perform more collisions 
with those particles coming from above, what leads to an increase of
pressure. An increase of pressure allows, in general, a strong
Coulomb friction and thus a sticking of the contact point. When 
pressure decreases, the particle surface does not longer stick on the 
wall.

Since the sticking might also occur between particles,
we examine the angular velocity of the particles in the
neighborhood of the sticking particle, i.e. the particles to the upper left
of particle \#371. In Fig.\ \ref{fig:fig3} we plot the angular
velocities of particles \#371, \#390, \#409, and \#428 for the simulations
(a) and (b). We observe from both figures as a response to 
friction, an auto-organization of the spins as also observed in 
1D experiments and simulations of rotating frictional cylinders
\cite{radjai95,radjai95b}. Spin stands here for the direction of angular
velocity of a particle. We see that the direct neighbor of \#371 to the
left and upwards, i.e. \#390, rotates in the opposite direction
as \#371. Thus a counterclockwise rotation of particle \#371 leads to a
clockwise rotation of \#390. This is consistent with the idea of 
friction reducing the relative surface velocity. The next
particle to the upper left, i.e. \#409, is again rotating couterclockwise,
following the same idea of frictional coupling. In Fig.\ \ref{fig:fig3}(a)
we observe only a weak coupling between \#409 and \#428, whereas in Fig.\ 
\ref{fig:fig3}(b) particle \#428 is also rotating clockwise, even when the
absolute value of the angular velocity is smaller. Thus we have not only a
stick-slip behavior of particles close to the side walls, but also a
coupling of the spins of neighboring particles. If the coupling is strong
enough, we observe an alternating, but decreasing angular velocity along a
line. We will return to this observation in subsection \ref{sec:order}. 

\subsection{Number of Collisions, Kinetic Energy and Pressure}

\label{sec:ncoll}

Since the occurrence of a crack is, in general, connected to a 
large number of collisions, we plot
in Fig.\ \ref{fig:fig4} the number of collisions per particle
per millisecond, $N_c$, for simulations (a) and (b). In Fig.\ \ref{fig:fig4}%
(a) we observe an increase in the number of collisions, which is related
to the first sticking event of particle \#371. However, particles deep
in the array possibly perform much more collisions, see \#428 in
Fig.\ \ref{fig:fig4}(a). Obviously, an increase in $%
N_c$ for one particle is connected to an increase in $N_c$ for the
neighbors. In Fig.\ \ref{fig:fig4}(b) we find values of $N_c$ comparable
for direct neighbors, i.e. \#371 and \#390.
At $t$ = 0.056 s we observe a drastic increase 
in $N_c$ which also involves the
particles deeper in the array, i.e. \#409 and \#428. 
\protect{\FFIGD}

In order to illustrate the reduced falling velocity, 
connected to a large number of
collisions we plot in Fig.\ \ref{fig:fig5} the averaged 
kinetic energy $K$ for particles between heights $z$ and $z+dz$
(we use here $dz$ = 0.002 m).
We plot $K = (1/N) \Sigma_{i=1}^N v_i^2$
(disregarding the mass of the particles), as a function
of the height, $z$. Note that $v_i$ is the velocity of one particle
relative to the walls, and not relative to the center of mass of
the falling sandpile. We observe different behavior for the simulations 
(a) and (b): A rather homogeneous deceleration of the pile in (a) and
one 'dynamic arch' connected to a strong deceleration in (b).
For strong friction at the walls (a), all particles in the system
are slowed down due to frequent collisions of the particles with the side
walls and inside the bulk. For the times $t$ = 0.05, 0.06, and 0.08 s we
observe, from the bottom indicated by the left vertical line in 
Fig.\ \ref{fig:fig5}, a decreasing $K$ with increasing height $z$ up to $%
z \approx 0.02$~m, where the slope of $K$ almost vanishes. In Fig.\ \ref
{fig:fig5}(b) the particles are falling faster at the beginning until
the first dynamic arch occurs at $t \approx 0.056$~s, 
see Fig.\ \ref{fig:fig1}(b). 
The high pressure exerted on the walls, together with the 
tangential friction at the walls, leads to a local deceleration, 
i.e. the dip in the $K$ profile for $t$ = 0.06 s. 
This dip identifies a dynamic arch of slow material
which temporarily blocks the flow. Later in time, the particles from above
arrive at the slower dynamic arch, which again leads to great pressure, a large
number of collisions and thus to a further reduction of $K$ (see the $K$
profile for $t$ = 0.08~s).
\protect{\FFIGE}

In order to understand how the pressure in the bulk is connected to $N_c$
and $K$ we plot in Fig.\ \ref{fig:fig6} the particle-particle pressure, $pp$,
in arbitrary units as a function of height for the simulations (a) and (b). 
$pp$ is here defined as the sum over the absolute normal part of momentum
change 
\begin{equation}
pp(z,t) = \Sigma | \Delta \vec P^{(n)} |,
\end{equation}
for each particle in a layer $[z,z+dz]$ in a time interval $[t,t-dt]$. For
this plot we use $dz = \sqrt{3} d$ (what corresponds to a height of two
particle layers) and the integration time is here $dt$ = 0.005 s. For
strong wall friction and low dissipation (a), 
we observe already at $t$ = 0.02~s a quite strong
pressure with a maximum close to the bottom of the pile. At $t$ = 0.05~s the
maximum in pressure moved upwards, not only within the falling pile but
also in coordinate $z$. Furthermore, the maximum is about one order of
magnitude larger than before. Later, the pressure peak moved further upwards
and at $t$ = 0.06 s begins to decrease until at larger times the array
is dilute and pressure almost vanished. For weak wall friction (b) 
we find only
a weak pressure at time $t$ = 0.02 s until at time $t$ = 0.06 s a
sudden increase of pressure, connected to the increase in $N_c$ and the
decrease in $K$, appears. The occurrence of two pressure peaks of
different amplitude (small pressure at the bottom and large pressure
at the top) is consistent with the predictions of Duran et al.
\cite{duran95}, which state that a small lower pile will
be decelerated less than a large upper pile. Thus the crack
continues opening.
\protect{\FFIGF}

\subsection{Long Range Rotational Order and Momentum Waves}

\label{sec:order}

We have learned from Figs.\ \ref{fig:fig2} and \ref{fig:fig3} that,
connected to a large number of collisions, the direction of the 
angular velocity, i.e. the spin of the particles, 
may be locally arranged in an alternating order along 
lines of large pressure. In
order to proof that this is not only a random event we plot in Fig.\ \ref
{fig:fig7}(a) some snapshots from simulation (a) and indicate clockwise and
counterclockwise rotation with black and white circles respectively. 
We observe, at
least in some parts of the system, that spins of the same direction are
arranged along lines. The spins of two neighboring lines have different
directions. The elongation of the ordered regions may be comparable
to the size of the system. Note, that
lines of equal spin are perpendicular to a line of strong pressure,
such that the order in Fig.\ \ref{fig:fig7}(a) indicates
an arch like structure. 

In Fig.\ \ref{fig:fig7}(b) we plot snapshots from simulation (b) and plot
the change of velocity, i.e. $\Delta v_x = v_x(t+\delta t) - v_x(t)$ and $%
\Delta v_z = v_z(t+\delta t) - v_z(t) + g~\delta t$ with $\delta t = 10^{-3}$%
~s. Comparing Figs.\ \ref{fig:fig1}(b) and \ref{fig:fig7}(b) we identify
the large number of collisions, starting from particle \#371, with a
momentum wave propagating from \#371 towards the left wall and also diagonally
upwards. When the momentum wave arrives at the left wall it is reflected and
moves mainly upwards. We relate this to the fact that the material below the
dynamic arch is falling faster than the dynamic arch, 
such that not much momentum change takes
place downwards. 
After several milliseconds the momentum wave is not longer
limited to some particles only, but has spread and builds now an active region
with great pressure, i.e. the dynamic arch. 

\FFIGG
Examining the rotational order in simulation (b), we
observe that the spin order is a consequence of the momentum wave. 
In our model, strong coupling is related to a large number of collisions 
and thus to a large pressure. This is due to the fact, that informations 
about the state of the particles are exchanged only on contact. 
Therefore, lines of equal spin are mostly perpendicular to
the lines of great pressure, a finding that is also discussed
in Ref.\cite{radjai95c}. 

\section{Summary and Conclusion}

\label{conclusion}

We presented simulations of a 2D granular model material, falling inside a
vertical container with parallel walls. We observe fractures in
the material which were described as a so-called `discontinuous 
decompaction' which is the result of many cracks breaking the
granular assembly into pieces from the bottom to the top.
The case of high friction and quite low dissipation (a) is a system which 
shows the same behavior as also found in various experiments. 
Many cracks occur and interfere. When we investigate an exemplary 
situation with rather low friction and high dissipation 
(b) we observe one isolated crack. We followed in detail the events
which lead to this crack. Due to
fluctuations of pressure (equivalent to fluctuations in
the number of collisions, $N_c$), some particles may transfer
a part of their linear momentum into rotational momentum. This
happens when the surface of a particle with small angular velocity
sticks on the wall. Sticking means here that the velocites of 
particle surface and wall surface are adapted.

The momentum wave, starting from such a particle, leads to a region
of large pressure, which spans the width of the system, i.e. a dynamic arch.
Due to the strong pressure, the dynamic arch is slowed down by friction with
the walls. The material coming from above hits the dynamic arch
such that a density and pressure wave, propagates upwards inside 
the system. 

Under conditions with quite strong wall friction and 
rather weak dissipation, the fluctuations in the
system and also the coupling with the walls are greater, such that many
particles are sticking on the walls. This leads to several pressure waves,
interferring with each other such that the system is slowed down 
more homogeneously.

With the dynamic simulations used, we were able to reproduce
the experimentally observed discontinuous decompaction \cite{duran95} 
and to propose an explanation, how the cracks occur. 
The open question remains, if the situation discussed here is
relevant for all types of experiments in restricted geometries. 
The features described here, i.e. rotational order, stick-slip behavior,
and momentum waves are not yet observed in experiments. 
Besides, the phenomenon of stress fluctuations
has been shown to be important for both, the behavior of static \cite{liu95}
and of quasistatic granular systems \cite{pouliquen95}. 
Furthermore, the behavior of cracks in polydisperse or three 
dimensional systems is still an open problem.

\section*{Acknowledgments}

We thank S. Roux for interesting discussions and gratefully acknowledge
the support of the European Community (Human Capital and Mobility) and
of the PROCOPE/APAPE scientific collaboration program. The
group is part of the French Groupement de Recherche sur la Mati\`{e}re
H\'{e}t\'{e}rog\`{e}ne et Complexe of the CNRS\ and of a CEE network HCM.




\end{document}